\documentclass[12pt]{article}
\usepackage{epsfig}
\usepackage{latexsym}
\usepackage{axodraw}

\begin{document}

\begin{titlepage}

\baselineskip 24pt

\begin{center}

{\Large {\bf Photo-Transmutation of Leptons}}\\

\vspace{.5cm}

\baselineskip 14pt

{\large Jos\'e BORDES}\\
jose.m.bordes\,@\,uv.es\\
{\it Departament Fisica Teorica, Universitat de Valencia,\\
  calle Dr. Moliner 50, E-46100 Burjassot (Valencia), Spain}\\
\vspace{.2cm}
{\large CHAN Hong-Mo}\\
chanhm\,@\,v2.rl.ac.uk \\
{\it Rutherford Appleton Laboratory,\\
  Chilton, Didcot, Oxon, OX11 0QX, United Kingdom}\\
\vspace{.2cm}
{\large Jacqueline FARIDANI}\\
jfarida\,@\,osfi-bsif.gc.ca\\
{\it Office of the Superintendent of Financial Institutions of
Canada,\\ 121 King St. W., Toronto, ON M5H 3T9, Canada}\\
\vspace{.2cm}
{\large TSOU Sheung Tsun}\\
tsou\,@\,maths.ox.ac.uk\\
{\it Mathematical Institute, University of Oxford,\\
  24-29 St. Giles', Oxford, OX1 3LB, United Kingdom}

\end{center}

\vspace{.3cm}

\begin{abstract}

By photo-transmutation of leptons we mean photon-lepton reactions of 
the following type: $\gamma l_\alpha \longrightarrow \gamma l_\beta$ 
with $l_\alpha \neq l_\beta$, occurring as a consequence of the lepton 
mass matrix changing its orientation (rotating) under changing scales.  
In this paper, we first discuss these reactions in general terms, then 
proceed to the calculation of their cross sections in two specific 
schemes, one within the framework of the conventional Standard Model, 
the other being the so-called Dualized Standard Model we ourselves 
advocate.  Although the cross section obtained is generally small the 
calculation reveals certain special circumstances where these reactions 
may be accessible to experiment, for example with virtual photons in 
LEP for $\gamma e \longrightarrow \gamma \tau$, and in BEPC for 
$\gamma e \longrightarrow \gamma \mu$. 

\end{abstract}

\end{titlepage}

\clearpage

\baselineskip 14pt

\setcounter{section}{0}
\setcounter{equation}{0}
\def\theequation{\arabic{section}.\arabic{equation}}

\section{Introduction}

By photo-transmutation of leptons, we mean the class of reactions:
\begin{equation}
\gamma + l_\alpha \longrightarrow \gamma + l_\beta,
\label{phototrans}
\end{equation}
with $l_\alpha$ and $l_\beta$ being two different charged lepton states,
which can occur in consequence of the lepton mass matrix changing its 
orientation in generations space (rotates) with changing energy scales
\cite{rge}.  
Such reactions, of course, do not occur when the lepton states $e, \mu$, 
or $\tau$ are, and remain, eigenstates of the lepton mass matrix.  But if 
the mass matrix rotates with changing scales, as is seen to be the case 
\cite{impromat} in the Standard Model when there is nontrivial mixing 
between up- and down-states \cite{ckm,mns}, then at any scale other
than that at which 
the leptons states are defined as eigenstates, the mass matrix will no 
longer be diagonal and transmutational reactions such as (\ref{phototrans}) 
with $\alpha \neq \beta$ can occur.  This is a special case of a general
class of fermion transmutation phenomena arising from rotating fermion
mass matrices, for a more detailed discussion of which the reader is 
referred to our companion paper \cite{impromat}.  

Given that in recent experiments on neutrino oscillations 
\cite{soudan,chooz,superk}, 
appreciable lepton mixing is indicated, it is already incumbent upon 
us, even when working within the present Standard Model framework, to 
investigate whether transmuational effects can be observed, although 
these are generally expected to be small.  Going beyond the Standard 
Model, however, especially in schemes aimed at explaining the generation 
puzzle such as the Dualized Standard Model that we ourselves advocate 
\cite{physcons,ourckm,ournuos,phenodsm,dualgen}, such an investigation 
becomes imperative, since there may then be other forces driving 
the rotation of the fermion mass matrix, leading to enhancements of 
transmutational effects beyond that given above by nontrivial mixing 
in the Standard Model.  It would thus be important to check for the 
consistency of such schemes whether the transmutational effects they 
predict would remain within the present experimental bounds and, if 
they do, whether they may be accessible to experimental tests in the 
not too distant future.

The photo-transmutation of leptons (\ref{phototrans}) is one of the 
simplest example we could find of transmutational effects to calculate, 
and also probably one of the most easily accessible to experimental 
scrutiny.  We propose therefore to examine this in some detail below 
both for general interest and with the ulterior motive of testing our 
Dualized Standard Model.     

\begin{figure}[ht]
\begin{center}
{\unitlength=1.0 pt \SetScale{1.0} \SetWidth{1.0}
\begin{picture}(350,100)(0,0) 
\Line(30,50)(100,50)
\Photon(30,50)(0,80){3}{5}
\Photon(100,50)(130,80){3}{5}
\Line(0,20)(30,50)
\Line(100,50)(130,20)

\Text(-10,20)[]{$l_\alpha$}
\Text(140,20)[]{$l_\beta$}
\Text(-10,80)[]{$\gamma$}
\Text(140,80)[]{$\gamma$}
\Text(25,30)[]{$p$}
\Text(105,30)[]{$p'$}
\Text(25,70)[]{$k$}
\Text(105,70)[]{$k'$}
\Text(65,0)[]{$(a)$}

\Line(230,50)(300,50)
\Photon(230,50)(320,80){-3}{10}
\Photon(300,50)(210,80){-3}{10}
\Line(200,20)(230,50)
\Line(300,50)(330,20)
\Text(190,20)[]{$l_\alpha$}
\Text(340,20)[]{$l_\beta$}
\Text(190,80)[]{$\gamma$}
\Text(340,80)[]{$\gamma$}
\Text(225,30)[]{$p$}
\Text(305,30)[]{$p'$}
\Text(228,68)[]{$k$}
\Text(303,68)[]{$k'$}
\Text(265,0)[]{$(b)$}
\end{picture} }
\end{center}
\caption{Compton-like diagrams for photo-transmutation of leptons}
\label{Comptdiag}
\end{figure}
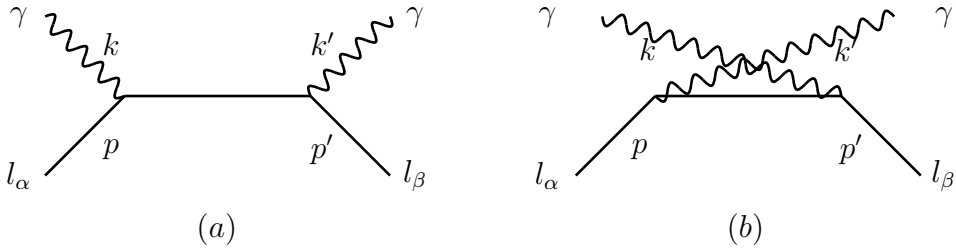

To leading order in perturbation theory, the cross section for the process
(\ref{phototrans}) is given by the Feynman diagrams in Figure \ref{Comptdiag}
which are formally the same as those for ordinary Compton scattering except 
that the lepton mass matrix pertaining to the lepton lines in the diagrams,
referring as it does to the energy scale $\sqrt{s}$ at which the experiment 
is performed, would in general be nondiagonal in the physical lepton states
$e, \mu$, and $\tau$ as identified in the incoming and outgoing lepton 
beams.  Because of this peculiarity, some care is needed in adapting the 
usual Feynman rules to this case.  

\setcounter{equation}{0}

\section{Kinematics}

To measure the cross section of a transmutation process, say e.g. $\gamma
\mu \longrightarrow \gamma e$, one would send in a beam of muons and install
downstream a detector for electrons.  However, in the
case that the lepton mass 
matrix rotates, as we have discussed in some detail in our companion paper 
\cite{impromat}, both what we call a muon and what we call an electron have
to be defined at some prescribed scales.  Two prescriptions were considered,
for details of which the reader is referred to that paper:
(i) ``fixed scale diagonalization'' (FSD) where the state vectors for the 
leptons of all 3 generations are defined at the same chosen scale, and
(ii) ``step-by-step diagonalization'' (SSD) where the state vectors for the 
3 leptons are each defined at its own mass scale.  The two prescriptions
differ in that the mass matrix is diagonal in (i) only at the chosen scale 
but in (ii) whenever the scale equals any of the 3 lepton masses.  In either 
case, however, since the mass matrix rotates, the state vectors thus defined 
have no reason to be eigenstates of the mass matrix at the scale of the 
scattering energy $\sqrt{s}$ at which the transmutation measurement is 
carried out.  This means that from the point of view of the experiment,
neither the incoming muon nor the outgoing electron has a definite mass.
Instead, each will be a linear combination of the states with definite 
masses at the scale of the scattering energy, namely the eigenstates of
the mass matrix at that scale with the eigenvalues $m_i$.

At the scattering scale, therefore, let us write the state vector of an
``external'' (i.e. incoming or outgoing) state $U_\alpha$ as:
\begin{equation}
U_\alpha = S_{\alpha i} U_i,
\label{Ualpha}
\end{equation}
where
\begin{equation}
U_1 = \left( \begin{array}{c} u_1 \\ 0 \\ 0 \end{array} \right); \ \ \
U_2 = \left( \begin{array}{c} 0 \\ u_2 \\ 0 \end{array} \right); \ \ \ 
U_3 = \left( \begin{array}{c} 0 \\ 0 \\ u_3 \end{array} \right),
\label{Ui}
\end{equation}
with $S_{\alpha i}$ being the rotation matrix relating the external state
$\alpha$ to the ``internal states'' $i$.  Each $u_i$, being a state with 
a definite mass $m_i$ at the scale under consideration, can be assigned a 
definite 4-momentum $p_i$ satisfying the mass condition $p_i^2 = m_i^2$
and be taken as a solution of the free Dirac equation:
\begin{equation}
(p\llap/_i - m_i) u_i(p_i) = 0.
\label{Diraceqi}
\end{equation}
The state $U_\alpha$ at the same scale, however, does not have a definite
mass, and therefore cannot be assigned a definite 4-momentum, although, 
of course, it has a mass $m_\alpha$ and a 4-momentum $p$ at the scale 
where it is defined as an eigenstate.  In trying to evaluate the Feynman 
diagrams (a) and (b) of Figure \ref{Comptdiag}, the question then arises 
how the momenta $p_i$ of the eigenstates $u_i(p_i)$ should be related 
first, to one another, and secondly, to the incoming (outgoing) momentum 
$p$ ($p'$) of the lepton states $l_\alpha$ ($l_\beta$).  

To answer this question, we recall a more familiar but parallel situation 
in neutrino oscillations.  A muon neutrino $\nu_\mu$ there, not being an 
eigenstate of the mass matrix, has also no definite 4-momentum, but can be 
expressed as a linear combination of the mass eigenstates $\nu_i, i = 1,2,3$, 
each of which can be assigned a definite 4-momentum $p_i$.  As to how these
$p_i$ are related to one another and to the given momentum of the $\nu_\mu$
beam, the answer depends on the problem being investigated.  In 
the case that the
incoming $\nu_\mu$ beam is given a definite 3-momentum ${\bf p}$ say, then 
the 3-momenta ${\bf p}_i$ by definition are all equal to ${\bf p}$, but the
energies $E_i = \sqrt{{\bf p}_i^2 + m_i^2}$ have to differ, and one has
neutrino oscillations in time as the beam propagates.  On the other hand,
in the case that
the incoming $\nu_\mu$ is given a definite energy, $E$ say, which 
is more likely in most experimental situations, then the energies $E_i$ 
are all equal to $E$, but the 3-momenta ${\bf p}_i$ differ, and one has 
neutrino oscillations in distance (i.e.\ the baseline).  Hence, by analogy
one sees that the answer to the question asked in the last paragraph will 
also depend on the problem addressed.  Since we have already specified that 
the diagrams (a) and (b) of Figure \ref{Comptdiag} are to be evaluated at 
the scale $\sqrt{s}$, we have already ascribed to these diagrams a definite 
$s$, namely that:
\begin{equation}
s = (p + k)^2 = (p_i + k)^2,
\label{identifys}
\end{equation}
common both to the incoming (outgoing) channels $\alpha$ ($\beta$) and to 
all eigen-channels $i$, which means that in this case, in contrast to the
above example in neutrino oscillations, neither the 3-momentum nor the 
energy, but only $s$ will have a definite value.

To evaluate the Feynman diagrams, we shall need the explicit relations 
between the different momenta $p_i$.  Given 4-momentum conservation:
\begin{equation}
p_i + k = p_i' +k',
\label{momencons}
\end{equation}
we note first that the 4-momentum transfer
\begin{equation}
t = (k - k')^2 = (p_i - p'_i)^2
\label{identifyt}
\end{equation} 
is also the same for all $i$, since $k$ and $k'$, being the photon momenta, 
have nothing to do with the rotation of the fermion states.  Expressing
then $p_j$ as a linear combination of $k, k'$ and $p_i$: 
\begin{equation}
p_j = a_j k + b_j k' + c_j p_i.
\label{pilincom}
\end{equation}
and imposing the conditions (\ref{identifys}) and:
\begin{equation}
p_i^2 = p_i'^2 = m_i^2,
\label{pipipmass}
\end{equation}
we obtain:
\begin{eqnarray}
c_j & = & \sqrt{\frac{(s - m_j^2)^2 +st}{(s - m_i^2)^2 + st}};
      \nonumber \\
b_j & = & \frac{c_j (s - m_i^2) - (s - m_j^2)}{t}; \nonumber \\
a_j & = & \frac{c_j (s + t - m_i^2) - (s + t - m_j^2)}{t}.
\label{cba}
\end{eqnarray}
Similarly, expressing $p_j'$ as:
\begin{equation}
p_j' = a_j' k' + b_j' k + c_j' p_i',
\label{pilincomp}
\end{equation}
we obtain:
\begin{equation}
c_j' = c_j; \ \ \ b_j' = b_j; \ \ \ a_j' = a_j.
\label{cbap}
\end{equation}
It is easy to check that the above relations between $p_i$'s and $p_i'$'s 
are both reflexive and transitive as they should be and that, in spite of 
appearance, are nonsingular at $t = 0$. 

Equivalently, the relations between the momenta derived in the preceding
paragraph may be expressed in terms of scalar products as follows:
\begin{eqnarray}
(p_i k) & = & \frac{1}{2} (s - m_i^2), \label{pik} \\
(p_i k') & = & \frac{1}{2} (s +t - m_i^2), \label{pikp} \\
(p_i p_j) & = & \frac{1}{t} \sqrt{[(s - m_i^2)(s + t - m_i^2) + m_i^2 t]
                [(s - m_j^2)(s + t -m_j^2) + m_j^2 t]} \nonumber \\
          &   & - \frac{1}{2 t} [(s - m_i^2)(s + t - m_j^2) 
                + (s - m_j^2)(s + t - m_i^2)], \label{pipj} \\
(p_i p_j') & = & \frac{1}{t} \sqrt{[(s - m_i^2)(s + t - m_i^2) + m_i^2 t]
                 [(s - m_j^2)(s + t -m_j^2) + m_j^2 t]} \nonumber \\ 
           &   & - \frac{1}{2 t} [(s + t - m_i^2)(s + t - m_j^2) 
                 + (s - m_j^2)(s - m_i^2)], \label{pipjp}
\end{eqnarray}
together with  some similar relations for $p_i'$, which may be convenient 
for invariant calculations in the future. 

\begin{figure}
\centerline{\psfig{figure=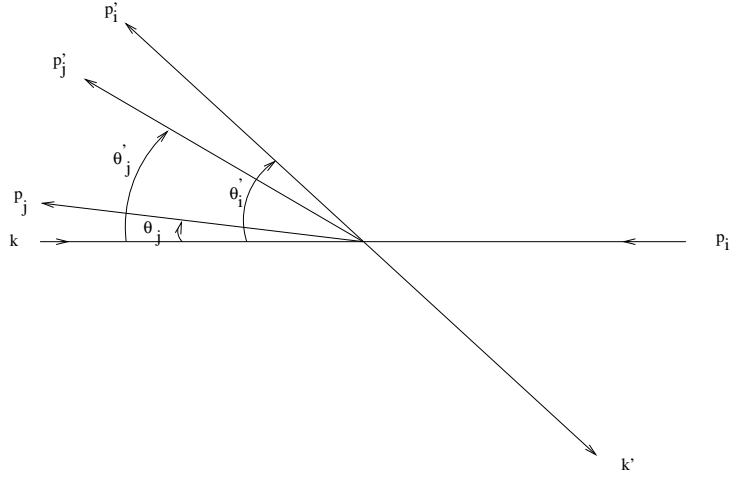,width=0.7\textwidth}}
\caption{The 3-momenta of the various particles in the $i$ cm frame}
\label{cmfig}
\end{figure}

However, for reasons which will be made clear later, we shall in this 
paper be working in a specific Lorentz frame, namely the cm frame of one 
chosen channel, say $i$, i.e. in the frame where the 3-momenta:
\begin{equation}
{\bf p}_i + {\bf k} = {\bf p'}_i + {\bf k'} = 0.
\label{cmframe}
\end{equation}
This does not correspond in general to the cm frame for another internal 
channel, say $j$, given the way that the internal momenta $p_i$ and $p_j$ 
are related.  Labelling in the chosen $i$ cm frame the angles and 3-momenta 
for the various particles as in Figure \ref{cmfig}, 
we can write their momenta as:
\begin{eqnarray}
k & = & (\omega_i, 0, 0, \omega_i), \nonumber \\
k' & = & (\omega_i, 0, \omega_i \sin \theta_i', \omega_i \cos \theta_i'), 
   \nonumber \\
p_i & = & (E_i, 0, 0, -\omega_i), \nonumber \\
p_i' & = & (E_i, 0, -\omega_i \sin \theta_i', -\omega_i \cos \theta_i'), 
   \nonumber \\
p_j & = & (E_j, 0, -\omega_j \sin \theta_j, -\omega_j \cos \theta_j),
    \nonumber \\ 
p_j' & = & (E_j, 0, -\omega_j \sin \theta_j', -\omega_j \cos \theta_j'),
\label{cmmomenta}
\end{eqnarray}
with
\begin{eqnarray}
E_i & = &  \frac{s + m_i^2}{2 \sqrt{s}}, \label{Eiinst}\\
\omega_i & = & \frac{s - m_i^2}{2 \sqrt{s}}, \label{omegaiinst}\\
\cos \theta_i' & = & 1 + \frac{2st}{(s - m_i^2)^2}. \label{cosipinst}
\end{eqnarray}
Hence, from (\ref{pilincom}), (\ref{pilincomp}), (\ref{cba}), and 
(\ref{cbap}), one easily obtains in terms of the invariants $s$ and $t$:
\begin{equation}
E_j = \frac{1}{2 \sqrt{s} t}[2 \sqrt{\{(s - m_j^2)^2 +st\}
   \{(s - m_i^2)^2 +st\}} - (s - m_i^2)\{2(s - m_j^2) + t\}], 
\label{Ej}
\end{equation}
with
\begin{equation}
\omega_j = \sqrt{E_j^2 - m_j^2}, 
\label{omegaj}
\end{equation}
and
\begin{equation}
\sin \theta_j = \frac{1}{\sqrt{s}} \frac{1}{\sqrt{-t}} \frac{1}
   {2 \omega_i \omega_j}[(s - m_i^2)\sqrt{(s - m_j^2)^2 + st} - (s - m_j^2)
   \sqrt{(s - m_i^2)^2 +st}], 
\label{sinj}
\end{equation}
\begin{equation}
\sin \theta_j' = \frac{1}{\sqrt{s}} \frac{1}{\sqrt{-t}} \frac{1}
   {2 \omega_i \omega_j}[(s - m_i^2)\sqrt{(s - m_j^2)^2 + st} - (s - m_j^2 +t)
   \sqrt{(s - m_i^2)^2 +st}], 
\label{sinjp}
\end{equation}
with
\begin{equation}
\theta_i' = \theta_j + \theta_j'.
\label{thetaeq}
\end{equation}
Again, in spite of appearance, these formulae are all nonsingular at $t =0$. 
They will be useful later for the calculation of the cross section.

\setcounter{equation}{0}

\section{The Invariant Amplitude}

The amplitude to leading order for the transmutation process (\ref{phototrans})
is then by (\ref{Ualpha}) given as:
\begin{equation}
{\cal M} = \sum_i S^*_{\beta i} {\cal M}_i S_{\alpha i},
\label{calM}
\end{equation}
where we have made use of the fact that the lepton mass matrix, and hence
also the scattering matrix, is diagonal in the (internal) states $i$.
Each ${\cal M}_i$ is a sum of two terms corresponding respectively to the 
two diagrams (a) and (b) of Figure \ref{Comptdiag}:
\begin{equation}
{\cal M}_i = {\cal M}_i^{(a)} + {\cal M}_i^{(b)},
\label{calMi}
\end{equation}
with
\begin{equation}
{\cal M}_i^{(a)} = -ie^2 \bar{u}(p'_i) \epsilon\llap/^*(k')
   \frac{(p\llap/_i + k\llap/) + m_i}{(p_i + k)^2 - m_i^2} 
   \epsilon\llap/(k) u(p_i),
\label{calMia}
\end{equation}
and
\begin{equation}
{\cal M}_i^{(b)} = -ie^2 \bar{u}(p'_i) \epsilon\llap/(k) 
   \frac{(p\llap/_i - k\llap/') + m_i}{(p_i - k')^2 - m_i^2} 
   \epsilon\llap/^*(k') u_i(p_i).
\label{calMib}
\end{equation}
Notice that although the last two formulae are formally the same as in 
ordinary Compton scattering, the quantities $m_i$ and $p_i$ which enter are
both dependent on the energy scale $\sqrt{s}$.

Given the rotation $S_{\alpha i}$ and the unusual kinemtics both of which
depend rather intricately on the energy scale, a check on some basic 
properties of the amplitude is warranted.  First, we wish to be assured
that the amplitude (\ref{calM}) vanishes at all scales if we put either 
$\epsilon(k) \rightarrow k$ or $\epsilon(k') \rightarrow k'$ as required
by gauge invariance.  This is indeed true for on putting $\epsilon(k) = k$ 
and using the equation of motion (\ref{Diraceqi}), one obtains for the 
contribution of the $s$-channel diagram (a) to ${\cal M}$ as:
\begin{equation}
{\cal M}^{(a)} \longrightarrow -ie^2 \sum_i S^*_{\beta i} \bar{u}(p'_i)
   \epsilon\llap/(k') u(p_i) S_{\alpha i},
\label{gaugeterma}
\end{equation}
but for the contribution of the $u$-channel diagram (b) as:
\begin{equation}
{\cal M}^{(b)} \longrightarrow ie^2 \sum_i S^*_{\beta i} \bar{u}(p'_i) 
   \epsilon\llap/(k') u(p_i) S_{\alpha i},
\label{gaugetermb}
\end{equation} 
so that the two terms do indeed cancel as required at any scale.  A similar
conclusion obtains also when we put $\epsilon(k') \rightarrow k'$.

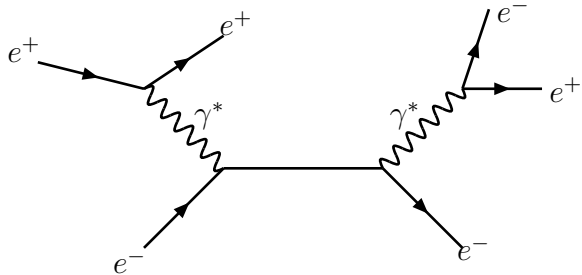
\begin{figure}
\begin{center}
{\unitlength=1.0 pt \SetScale{1.0} \SetWidth{1.0}
\begin{picture}(250,100)(0,0)

\Line(100,30)(160,30)
\Photon(160,30)(190,60){3}{6}
\ArrowLine(160,30)(190,0)
\ArrowLine(190,60)(200,90)
\ArrowLine(190,60)(220,60)
\Photon(100,30)(70,60){3}{6}
\ArrowLine(30,70)(70,60)
\ArrowLine(70,60)(100,80)
\ArrowLine(70,0)(100,30)
\Text(169,50)[]{$\gamma^*$}
\Text(210,90)[]{$e^-$}
\Text(230,60)[]{$e^+$}
\Text(195,0)[]{$e^-$}
\Text(95,50)[]{$\gamma^*$}
\Text(105,85)[]{$e^+$}
\Text(25,75)[]{$e^+$}
\Text(65,-5)[]{$e^-$}
\end{picture} }
\end{center}
\caption{Feynman diagram for the reaction $e^+ e^- \longrightarrow e^+ e^- 
e^+ e^-$}
\label{pairprod}
\end{figure}

Next, we wish to check that the amplitude ${\cal M}$ has sensible pole
structures.  Consider first the $s$-channel poles which according to 
(\ref{calMia}) can occur at $(p_i + k)^2 = m_i^2$, or equivalently by 
(\ref{identifys}) at $s = m_i^2$, when the denominator vanishes.  For
the actual reaction $\gamma l_i \longrightarrow \gamma l_i$ with real
photons, however, $s = m_i^2$ corresponds to the reaction threshold where
the numerator in the amplitude also vanishes so that the pole is cancelled,
corresponding to the fact that the Compton cross section is finite in the 
Thomson limit.  But if the photon is taken off mass-shell, then the
cancellation need not work and a pole can occur.  Consider first the
simpler situation at energies below the $\tau$ mass where the mass matrix
is only 2-dimensional, with the higher of its two eigenvalues, $m_2$ say, 
coinciding by definition with $m_\mu$ as $\sqrt{s} \rightarrow \mu$.  The 
amplitude $\gamma^* l_2 \longrightarrow \gamma^* l_2$ will thus develop 
a pole at $s = m_\mu^2$, but the amplitude for the other internal channel
$\gamma^* l_3 \longrightarrow \gamma^* l_3$, of course, will not have 
this pole.  Consider now, however, the amplitudes for the processes 
$\gamma^* \mu \longrightarrow \gamma^* \mu$ and $\gamma^* e \longrightarrow 
\gamma^* e$ with amplitudes as given by (\ref{calM}).  Their pole structure
will differ depending on whether the lepton states are defined by the
FSD or the SSD prescription.  For the SSD prescription, the mass matrix
is by construction diagonal at the scale of $m_\mu$ so that the pole at
$s = m_\mu^2$ will occur in the amplitude for $\gamma^* \mu \longrightarrow 
\gamma^* \mu$ but not in that for $\gamma^* e \longrightarrow \gamma^* e$,
which is as it should be.  But for the FSD prescription where the mass
matrix at the scale $m_\mu$ need not be diagonal, both these amplitudes 
will generally develop a pole at $s = m_\mu^2$.  This means that for a process 
such as $e^+ e^- \longrightarrow e^+ e^- e^+ e^-$ as indicated in Figure
\ref{pairprod}, the cross section will show a sharp resonance peak at
$s = m_\mu^2$.  The physical meaning of this is clear, namely first a 
formation of a $\mu$ resonance from $\gamma^* e^-$ followed by its decay
into $e^- e^+ e^-$.  In other words, the apparent FSD anomaly here of 
the $\mu$ pole appearing in the ``wrong'' channel is the same as that 
pointed out in \cite{impromat} of the occurrence of lepton flavour-violating 
$\mu \longrightarrow e^- e^+ e^-$ decay.  A similar conclusion applies to
the pole at $s = m_\tau^2$ occurring in $\gamma^* l_1 \longrightarrow 
\gamma^* l_1$.  For SSD, the now $3 \times 3$ mass matrix is again by 
construction diagonal at $s = m_\tau^2$, so that this pole will not appear in 
any of the four processes $\gamma^* e(\mu) \longrightarrow \gamma^* e(\mu)$.  
But for FSD, this is no longer guaranteed.  We conclude, therefore, that 
with the SSD prescription for defining fermion states, which we ourselves 
prefer, the amplitude (\ref{calM}) has indeed throughout a reasonable 
$s$-channel pole structure.  But for the FSD case, there is the probable 
difficulty of lepton poles showing up in wrong channels, unless the 
prescription is supplemented by conditions making certain rotation matrix 
elements vanish at the pole positions.  One possibility would be to choose 
to diagonalize the mass matrix at a scale above all the lepton masses and 
to assume that the mass matrix stops rotating below the scale at which it 
is diagonalized, which we believe is what is often tacitly assumed in the 
literature.  In what follows, we shall take for granted some such assumption 
has been made for the FSD case although we are unclear whether this is in 
fact justified.  

Next, we examine the $u$-channel pole structure.  According to (\ref{calM})
and (\ref{calMib}), the amplitude has poles at:
\begin{equation}
u_i = (p_i - k')^2 = m_i^2,
\label{upoles}
\end{equation}
where we note that, $s$ and $t$ being constrained to be the same for all
$i$, 
\begin{equation}
u_i = 2 m_i^2 - s - t
\label{ui}
\end{equation}
will generally have different values for different $i$.  Given that in
channel $i$, the minimun value of $t$ in the physical range is:
\begin{equation}
t_{min \ i} = - (s - m_i^2)^2/s,
\label{tmini}
\end{equation}
the maximum value of $u_i$ in the physical range of the $i$ channel is:
\begin{equation}
u_{i\ max} = m_i^4/s,
\label{uimaxi}
\end{equation}
so that for $m_i \neq 0$ the pole in (\ref{upoles}) lies always outside 
the physical region as it should.  For $m_i = 0$, which is a case we 
shall frequently encounter when calculating later transmutation in the 
Dualized Standard Model scheme, the pole lies at $u_i = 0$, right at the 
edge of the physical range of $u_i$ in the $i$ channel.  However, this 
value of $u_i$ is not accessible within the physical range of the actual 
transmutation process for which the minimum value of $t$ is:
\begin{equation}
t_{min} = - (s - m_\alpha^2)(s - m_\beta^2)/s,
\label{tmin}
\end{equation}
giving, for $m_i = 0$:
\begin{equation}
u_{i\ max} = - m_\alpha^2 - m_\beta^2 + m_\alpha^2 m_\beta^2/s,
\label{uimax}
\end{equation}
which is necessarily negative and therefore a finite distance from the
pole.  We conclude therefore that the amplitude (\ref{calM}) will also 
not be troubled by $u$-channel poles getting into the physical region
whether with the FSD or with the SSD prescription.

We note, of course, that at energy scales different from the actual lepton 
masses, the poles in the amplitude would have shifted to noncanonical 
positions by virtue of the running mass eigenvalues.  But this would mean 
only slightly different energy dependences of the cross sections from 
conventional QED which would not be easily detectable.  Nor is this a 
new phenomenon special to the rotating mass matrix but one common to all 
theories with running masses, and need not therefore be further scrutinized 
at present.   

Having then passed these tests, the amplitude (\ref{calM}) will now be 
used to calculate cross sections for transmutational processes.

\setcounter{equation}{0}

\section{Amplitudes with Definite Spins and Polarizations}

At the first instance we shall investigate the differential spin- and 
polarization-summed cross section for the transmutational processes in 
(\ref{phototrans}).  Very elegant invariant methods reducing the spin
and polarization sums to traces of $\gamma$ matrices have of course been 
developed for calculating these type of cross sections.  However, these are 
not immediately applicable to the amplitude (\ref{calM}) here because in
taking the absolute value squared of the amplitude there will be crossed
terms involving factors such as $u_i(p_i) \bar{u}_j(p_j)$ the summing of
which over spin does not give a simple projection operator into positive
energy states as in the usual case.  Still, of course, the invariant
method can be adapted for application to the present rotated amplitude,
which indeed has been tried, but this would introduce many more $\gamma$
matrices and make the calculation of their traces unwieldily complicated.

Mainly for this practical reason, we have chosen to calculate instead 
the spin- and polarization-summed cross section for (\ref{calM}) by the
pedestrian method using a specific Lorentz frame and representation of 
the $\gamma$ matrices, and then summing the contributions of the various 
spin and polarization amplitudes.  Though less elegant, this has the
virtue of being physically more transparent and offers in the future, 
when conditions are ripe, the means for examining the cross sections
for the different spin and polarization components.

We choose to work in the cm frame of one of the channels $i$ for which 
the kinematics have already been worked out in a previous section, and
adopt for $\gamma$ matrices the Pauli-Dirac representation, namely:
\begin{equation}
\gamma_0 = \left( \begin{array}{cc} 1 & 0 \\
                                    0 & -1   \end{array} \right); \ \ \ 
\gamma_k = \left( \begin{array}{cc} 0 & \sigma_k \\
                                    -\sigma_k & 0   \end{array} \right),
\label{gammarep}
\end{equation}
where $\sigma_k$ are the standard Pauli matrices.  We quantize the spin 
of all the incoming leptons $j$ along the direction ${\bf p}_i$ and of 
all the outgoing leptons $j'$ along the direction ${\bf p}_i'$, while 
the photon polarizations we quantize along respectively ${\bf k}$ and 
${\bf k}'$.  This gives for the photon polarization vectors:
\begin{equation}
\epsilon\llap/_+(k) = \left( \begin{array}{cccc} 0 & 0 & 0 & \sqrt{2} \\
                                                 0 & 0 & 0 & 0 \\
                                                 0 & -\sqrt{2} & 0 &0 \\
                                                 0 & 0 & 0 & 0 \end{array}
   \right), \ \ \ 
\epsilon\llap/_-(k) = \left( \begin{array}{cccc} 0 & 0 & 0 & 0 \\
                                                 0 & 0 & -\sqrt{2} & 0 \\
                                                 0 & 0 & 0 &0 \\
                                                 \sqrt{2} & 0 & 0 & 0 
   \end{array} \right);
\label{epslash}
\end{equation}
and
\begin{eqnarray}
\epsilon\llap/_+(k') & = & \frac{1}{\sqrt{2}} \left( \begin{array}{cccc}
   0 & 0 & -i \sin \theta_i' & 1 + \cos \theta_i' \\
   0 & 0 & 1 - \cos \theta_i' & i \sin \theta_i' \\
   i \sin \theta_i' & -(1 + \cos \theta_i') & 0 & 0 \\
   -(1 - \cos \theta_i') & - i \sin \theta_i' & 0 & 0 \end{array} \right), 
   \nonumber \\
\epsilon\llap/_-(k') & = & \frac{1}{\sqrt{2}} \left( \begin{array}{cccc}
   0 & 0 & -i \sin \theta_i' & -(1 - \cos \theta_i') \\
   0 & 0 & -(1 + \cos \theta_i') & i \sin \theta_i' \\
   i \sin \theta_i' & 1 - \cos \theta_i' & 0 & 0 \\
   1 + \cos \theta_i' & - i \sin \theta_i' & 0 & 0 \end{array} 
   \right); \nonumber \\
   &&
\label{epslashp}
\end{eqnarray}
and for the lepton wave functions with definite spins:
\begin{eqnarray}
u_+(p_j) &=& \frac{1}{\sqrt{2(E_j + m_j)}} \left( \begin{array}{c}
   0 \\ E_j + m_j \\ 0 \\ \omega_j e^{- i \theta_j} \end{array}
   \right), \nonumber \\ 
u_-(p_j) &=& \frac{1}{\sqrt{2(E_j + m_j)}} \left( \begin{array}{c}
   E_j + m_j \\ 0 \\ -\omega_j e^{- i \theta_j} \\ 0 \end{array}
   \right);
\label{uspinpj}
\end{eqnarray}
and:
\begin{eqnarray}
u_+(p_j') & = & \frac{1}{2\sqrt{2(E_j + m_j)}} \left( \begin{array}{c}
  (E_j + m_j) (1 - e^{-i \theta_i'}) \\ (E_j + m_j) (1 + e^{-i \theta_i'}) \\
  \omega_j (e^{i \theta_j} - e^{-i \theta_j'}) \\
  \omega_j (e^{i \theta_j} + e^{-i \theta_j'}) \end{array} \right), \\ 
u_-(p_j') & = & \frac{1}{2\sqrt{2(E_j + m_j)}} \left( \begin{array}{c}
  (E_j + m_j) (1 + e^{-i \theta_i'}) \\ (E_j + m_j) (1 - e^{-i \theta_i'}) \\
  -\omega_j (e^{i \theta_j} + e^{-i \theta_j'}) \\
  -\omega_j (e^{i \theta_j} - e^{-i \theta_j'}) \end{array} \right).
\label{uspinpjp}
\end{eqnarray}
In these formulae, we follow the near standard convention, say for example
in \cite{Mandshaw}, apart from the removal of a factor $1/\sqrt{m_j}$ in
the normalization of $u(p)$ which is more convenient for our case where
$m_j$ can be zero.  

With the photon polarization vectors in (\ref{epslash}) and (\ref{epslashp})
and the lepton wave functions in (\ref{uspinpj}) and (\ref{uspinpjp}), it 
is straightforward to calculate the various spin and polarization amplitudes.
We obtain the following, where superscripts denote the photon polarizations
and subscripts the lepton spins, while right-hand indices correspond to
incoming and left-hand indices outgoing states.  By parity, amplitudes 
are unchanged with the signs of all superscripts and subscripts reversed,
and by angular momentum conservation, only two amplitudes are nonzero for 
the $s$-channel diagram (a).
\begin{eqnarray}
({\cal M}^{(a)})^{++}_{++} & = & \frac{-i e^2}{2} \frac{1}{(s - m_j^2)}
   \{ \omega_i [E_j + m_j + 2 \omega_j e^{-i \theta_j} \nonumber \\
   && {} + (E_j - m_j)
   e^{-2i \theta_j}] (1 + e^{i \theta_i'}) 
   + 2 i \omega_j^2 \sin \theta_j (e^{i \theta_j'} - e^{-i
   \theta_j}) \}; \nonumber \\
({\cal M}^{(a)})^{-+}_{-+} & = & \frac{-i e^2}{2} \frac{1}{(s - m_j^2)}
   \{ \omega_i [- E_j - m_j + (E_j - m_j) e^{-2i \theta_j}](1 - 
   e^{i\theta_i'}) \nonumber \\ 
   & & {} + 2i \omega_j^2 \sin \theta_j (e^{i \theta_j'} - e^{-i 
   \theta_j}) \}; \nonumber \\
({\cal M}^{(b)})^{++}_{++} & = & \frac{- e^2}{4} \frac{(1 - e^{i \theta_i'})}
   {(u_j - m_j^2)} \times \nonumber \\
   && \{ 2 \omega_j e^{-i \theta_j} [\omega_j \sin \theta_j
   (1 - \cos \theta_i') + \omega_i \sin \theta_i' + \omega_j \cos \theta_j
   \sin \theta_i'] \nonumber \\
   & & {} + [(\omega_j^2 - \omega_i E_j + \omega_i m_j) e^{-2i \theta_j} 
   + (\omega_j^2 - \omega_i E_j - \omega_i m_j)]\sin \theta_i'\};
   \nonumber \\
({\cal M}^{(b)})^{-+}_{-+} & = & \frac{e^2}{4} \frac{(1 + e^{i \theta_i'})}
   {(u_j - m_j^2)} \times \nonumber \\
   &&\{ [\omega_j^2 - \omega_i E_j + \omega_i m_j] e^{-2i \theta_j}
   - [\omega_j^2 - \omega_i E_j - \omega_i m_j] \} \sin \theta_i';
   \nonumber \\
({\cal M}^{(b)})^{-+}_{++} & = & \frac{e^2}{4} \frac{(1 - e^{i \theta_i'})}
   {(u_j - m_j^2)} \times \nonumber \\
   &&\{ 2 \omega_j e^{-i \theta_j} [\omega_j \sin \theta_j
   (1 + \cos \theta_i') + \omega_i \sin \theta_i' - \omega_j \cos \theta_j
   \sin \theta_i'] \nonumber \\
   & & - [(\omega_j^2 - \omega_i E_j + \omega_i m_j) e^{-2i \theta_j} 
   + (\omega_j^2 - \omega_i E_j - \omega_i m_j)]\sin \theta_i'\};
   \nonumber \\   
({\cal M}^{(b)})^{++}_{-+} & = & \frac{e^2}{4} \frac{(1 + e^{i \theta_i'})}
   {(u_j - m_j^2)} \times \nonumber \\
   &&\{ [\omega_j^2 - \omega_i E_j + \omega_i m_j]
   e^{-2i \theta_j}
   - [\omega_j^2 - \omega_i E_j - \omega_i m_j] \} \sin \theta_i';
   \nonumber \\
({\cal M}^{(b)})^{--}_{++} & = & \frac{-i e^2}{4} \frac{(1 + e^{i \theta_i'})}
   {(u_j - m_j^2)} \times \nonumber \\
   &&\{ - 2 \omega_j e^{-i \theta_j} [(\omega_j \cos \theta_j
   + \omega_i) (1 + \cos \theta_i') + \omega_j \sin \theta_j \sin \theta_i']
   \nonumber \\
   & & {} + [(\omega_j^2 - \omega_i E_j + \omega_i m_j) e^{-2i \theta_j} 
   + (\omega_j^2 - \omega_i E_j - \omega_i m_j)] (1 + \cos \theta_i') \};
   \nonumber \\ 
({\cal M}^{(b)})^{+-}_{++} & = & \frac{-i e^2}{4} \frac{(1 + e^{i \theta_i'})}
   {(u_j - m_j^2)} \times \nonumber \\
   &&\{ 2 \omega_j e^{-i \theta_j} [(\omega_j \cos \theta_j
   - \omega_i) (1 - \cos \theta_i') - \omega_j \sin \theta_j \sin \theta_i']
   \nonumber \\
   & & {} - [(\omega_j^2 - \omega_i E_j + \omega_i m_j) e^{-2i \theta_j} 
   + (\omega_j^2 - \omega_i E_j - \omega_i m_j)] (1 - \cos \theta_i') \};
   \nonumber \\ 
({\cal M}^{(b)})^{+-}_{-+} & = & \frac{-i e^2}{4} \frac{(1 - e^{i \theta_i'})}
   {(u_j - m_j^2)} \times \nonumber \\
   &&\{ [\omega_j^2 - \omega_i E_j + \omega_i m_j] e^{-2i \theta_j}
   - [\omega_j^2 - \omega_i E_j - \omega_i m_j] \} (1 - \cos \theta_i');
   \nonumber \\
({\cal M}^{(b)})^{--}_{-+} & = & \frac{-i e^2}{4} \frac{(1 - e^{i \theta_i'})}
   {(u_j - m_j^2)} \times \nonumber \\
   &&\{-[\omega_j^2 - \omega_i E_j + \omega_i m_j] 
   e^{-2i \theta_j}
   + [\omega_j^2 - \omega_i E_j - \omega_i m_j] \} (1 + \cos \theta_i').
   \nonumber \\
\label{amplis}
\end{eqnarray}
For $j = i$, these amplitudes reduce to the amplitudes for ordinary Compton
scattering, affording thus a check of the above expressions.  

With these formulae, given the rotation matrix $S_{\alpha i}$, one can 
then calculate by summing over $j$ the amplitudes (\ref{calM}) of any 
transmutation process $\gamma l_\alpha \longrightarrow \gamma l_\beta$
for any spin and polarization combinations.

\setcounter{equation}{0}

\section{The Rotation Matrix}

Up to the present section, our formalism has been developed generally 
for any rotating mass matrix, without reference to a particular scheme 
for how the rotation is generated.  However, as can be seen in (\ref{calM}),
(\ref{calMia}) and (\ref{calMib}), the transmutation amplitude depends on 
the rotation matrix elements $S_{\alpha i}$ and the mass eigenvalues $m_i$,
all of which are functions of the energy scale $\sqrt{s}$ and depend on 
the scheme or model from which the rotating mass matrix is derived, which 
has thus now to be specified.  In our companion paper \cite{impromat}, 
two explicit schemes at more or less opposite ends of a spectrum of 
possibilities were considered, which are also what we have chosen to 
investigate in detail in this paper.  One is the so-called NSM scheme 
where the rotation of the fermion mass matrix is driven just by the 
nondiagonal lepton mixing or MNS matrix \cite{mns}
via the standard renormalization 
group equations (1.1) and (1.2) in \cite{impromat} of the (conventional) 
Standard Model.  Lepton states are here defined by the FSD prescription.  
The other is the so-called DSM (Dualized Standard Model) scheme where 
the rotation of the mass matrix is driven by new generation-changing 
forces not normally considered in the Standard Model.  In this latter 
scheme, even the mixing matrices of quarks and leptons are themselves 
consequences of the mass matrix rotation and even calculable, 
with the results 
so far obtained already giving a quite cogent explanation for the 
characteristic quark and lepton mixing patterns seen in experiment.  (See 
e.g. \cite{dualgen} for a summary.)  In this DSM scheme, specification of 
fermion states by the SSD prescription is essential.

Since these two schemes have already been described in our companion paper
\cite{impromat} with further details given elsewhere in the literature,
we shall note here only those points which are of particular relevance
to the present calculation.  Given that in the NSM scheme, the mass matrix
rotation is driven by the nondiagonal mixing matrix, while in the DSM
case, the nondiagonal mixing matrix itself results from the mass matrix
rotation, it follows that the transmutation effects arising from the 
rotation will in general be stronger in the DSM than in the NSM scheme.
(An exception is the pole effect in ``wrong channels'' for the NSM scheme,
because of the FSD prescription adopted, as already explained in section 3.) 
And since, as we shall see, transmutation effects are usually quite 
small, although some can
be near the present limit of experimental detectability, it 
is the DSM scheme which will be examined here in greater detail, being 
potentially of more immediate experimental interest. 

\begin{figure}
\centerline{\psfig{figure=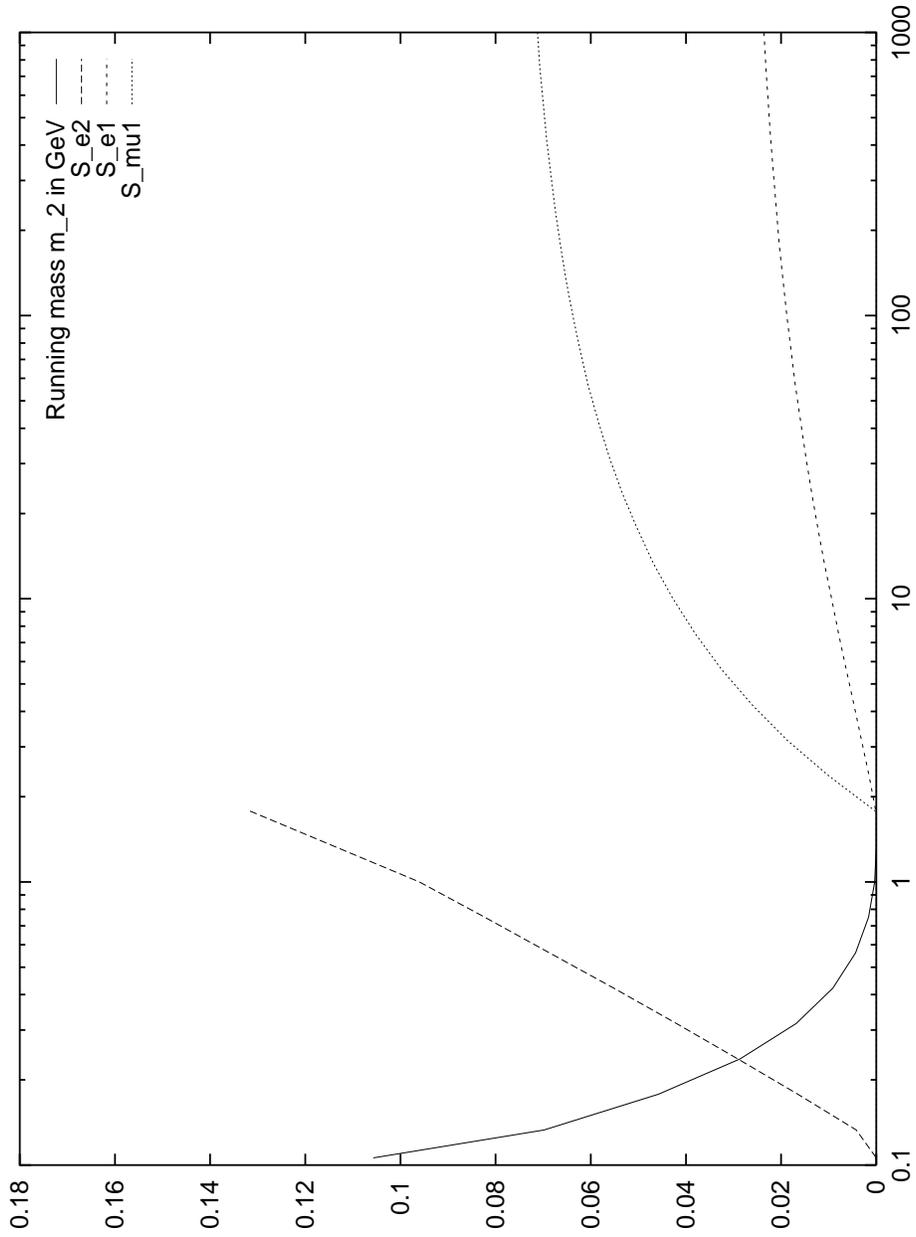,width=0.88\textwidth}}
\caption{The rotation matrix $S_{\alpha i}$ and the mass eigenvalue 
$m_2$ given as functions of the energy scale.  By construction, the 
scheme has $m_1 = m_\tau, m_2 = 0$ at any scale above the $\tau$ mass, 
and $m_3 = 0$ at any scale above the $\mu$ mass.}
\label{Salphai}
\end{figure}

As explained in \cite{impromat}, with its 3 free parameters already 
fitted to the mass ratios $m_c/m_t, m_\mu/m_\tau$ and the Cabibbo angle 
\cite{phenodsm}, the DSM scheme gives a parameter-free prediction for 
the rotating mass matrices as a function of the energy scale, which
in the case of charged leptons is found in Figure 3 of \cite{impromat}. 
As a result, the mass eigenvalues $m_i$ as well as the rotation matrices 
$S_{\alpha i}$ from the physical states $\alpha$ to the mass eigenstates 
$i$ are also known at any scale.  We have thus all the information needed 
to carry out the above calculation of transmutation cross sections.  
Although the accuracy and validity of these predictions are conditioned 
by the present limitations of the scheme and subject to possible amendments 
in the future when understanding improves, they will be sufficient for the 
exploratory purposes of this paper.  For ease of 
reference\footnote{We adopt here the convention of our earlier papers 
on the DSM
in labelling the fermion eigenstates in order of their mass eigenvalues,
i.e. with the heaviest labelled as 1, the second as 2, and the lightest 
as 3, which is more natural in this scheme.  Unfortunately, this is the 
reverse of the the standard convention for labelling neutrinos adopted 
in \cite{impromat} which follows the historical order of their
discovery.}, the values of 
$m_i$ and $S_{\alpha i}$ are given explicitly in Figure \ref{Salphai}.

For the NSM scheme with which we shall do no explicit detailed calculation, 
we need only note that the mass eigenvalues $m_i$, like the physical lepton
masses, are hierarchical, meaning $m_1 \gg m_2 \gg m_3$, while the rotation 
matrix $S_{\alpha i}$, like the mass matrix itself, have non-diagonal 
elements which are zero at the fixed scale chosen for diagonalization and 
then increase approxmiately linearly with the logarithm of the energy scale.
Their actual values can be calculated from eq. (4.2) of \cite{impromat}.

The mass eigenvalues $m_i$ enter in the diagonal amplitudes ${\cal M}_i$, 
and the rotation matrix $S_{\alpha i}$ in evaluating the transmutation 
amplitude with ({\ref{calM}) where a sum over the eigenstates $i$ has to 
be performed.  In performing this sum, two points of detail are noteworthy.  
First, in the DSM case, the mass matrix at any scale being a factorizable 
matrix has always only one non-zero eigenvalue, say $m_h$ which is $m_1$ 
for energies above the $\tau$ threshold but $m_2$ below.  The amplitudes 
${\cal M}_i$ above the $\tau$ threshold for $i = 2, 3$ are thus identical.  
Hence, by the unitarity of $S_{\alpha i}$, the sum over $i$ in (\ref{calM}) 
for the transmutation amplitude takes always the following simple form for 
$\alpha \neq \beta$:
\begin{equation}
\sum_{i} S_{\alpha i} {\cal M}_{i} S_{\beta i}
   = S_{\alpha h} S_{\beta h} [{\cal M}_{h} - {\cal M}_{3}].
\label{sumoveri}
\end{equation}
Therefore, if the difference ${\cal M}_{h} - {\cal M}_{3}$ is small 
compared with the amplitudes themselves, which it often will be, then 
it is much better to use the latter formula than to do the sum over $i$ 
directly, for the sum will involve large cancellations, requiring 
thus knowing $S_{\alpha i}$ to a greater accuracy than is warranted by
the DSM scheme in its present form.  Also in the NSM scheme, the same
statement will apply so long as the fermion mass spectrum is hierarchical
for in comparison to the heaviest generation, the two lighter ones will
appear degenerate so that the last formula in (\ref{sumoveri}) will be
a very good approximation.

Secondly, the last formula in (\ref{sumoveri}) also offers a link-up of
our present calculation to the estimates made in our companion paper
\cite{impromat}.  To be specific, consider the $s$-channel diagram.
The amplitudes ${\cal M}_{h}$ and ${\cal M}_{3}$ differ only in the
masses $m_h$ and $m_3$ where $m_3 \sim 0$.  Thus, if $m_h$ is small
compared with the energy $\sqrt{s}$ as is often the case, then the
difference between the two amplitudes is of order $m_h/\sqrt{s}$ times
the amplitude itself.  Noting next that $S_{\alpha h} S_{\beta h} m_h
= \sum_i S_{\alpha i} S_{\beta i} m_i = \langle \alpha|m|\beta \rangle$,
exactly in DSM because $m_i = 0, i \neq h$, and approximately in NSM so 
long as the spectrum is hierarchical, we conclude that the transmutation
amplitude is suppressed compared with that for diagonal processes by a 
factor $\langle \alpha|m|\beta \rangle/\sqrt{s}$, precisely as suggested 
in \cite{impromat}.

As we shall see, the formula (\ref{sumoveri}) gives also some useful
relations between cross sections of different transmutation processes.

\setcounter{equation}{0}

\section{Spin- and Polarization-Summed Cross Sections}

In calculating the cross sections of the various transmutation processes 
in (\ref{phototrans}), we find it easiest to work in the cm frame of the 
channel with the lowest lepton mass eigenvalue $m_3$, which in the DSM 
scheme is always zero.  Hence in the chosen frame, only the amplitudes 
for the massive lepton channel $h$ are complicated, while those of 
the remaining channel(s) reduce just to the cm amplitudes of Compton 
scattering in the limit $m_e \rightarrow 0$ and take particularly simple 
forms.  Applying then the formula (\ref{sumoveri}) to each of the amplitudes 
in (\ref{amplis}), summing over all 3 channels above the $\tau$-threshold
but over only channels 2 and 3 below the $\tau$-threshold, we obtain the 
8 transmutation amplitudes each with a definite combination of polarizations 
and spins.  Taking next the absolute values squared of these amplitudes 
and summing over all 8 spin and polarization combinations one obtains, 
for energies measured in GeV, the desired cross section as:
\begin{equation}
\frac{d \sigma}{d \Omega} = \frac{1}{16 \pi^2} \frac{\omega'}{\omega}
   \frac{1}{4 s} \sum_{spin} \sum_{pol} |{\cal M}|^2 \times 0.3894 
   \ \ {\rm mb/sr}.
\label{diffxsec}
\end{equation}
Given that all the variables appearing in (\ref{amplis}) have already been
expressed in terms of the invariants $s$ and $t$ in section 2, we can now 
evaluate the spin- and polarization-summed cross section as functions of 
$s$ and $t$ which, we recall, have the same meaning for all the internal 
channels $i$ as well as for the actual transmutation process $\gamma l_\alpha 
\longrightarrow \gamma l_\beta$.

Following this procedure, we have calculated explicitly the differential 
cross sections for the following 4 transmutation processes: $\gamma e 
\longrightarrow \gamma \mu (\tau)$ and $\gamma \mu \longrightarrow \gamma 
e (\tau)$, ignoring for the moment $\tau$ initiated processes since no 
$\tau$ target or beam is likely to be available in the foreseeable future.  
Of the 4 processes calculated, the cross sections of $\gamma e \longrightarrow 
\gamma \mu$ and $\gamma \mu \longrightarrow \gamma e$ are almost identical 
for the same values of $s$ and $t$, as can be easily understood.  In all 4
cases, as in ordinary Compton scattering, the cross sections are dominated 
at most energies by the $u$-channel pole(s).  However, we have found it 
more convenient to plot the cross sections not as functions of $u$ but as 
functions of the variable $s + t$ which is close to $u$ but, in contrast 
to $u$, has a common meaning for all channels, both internal and external.  
\begin{figure}
\centerline{\psfig{figure=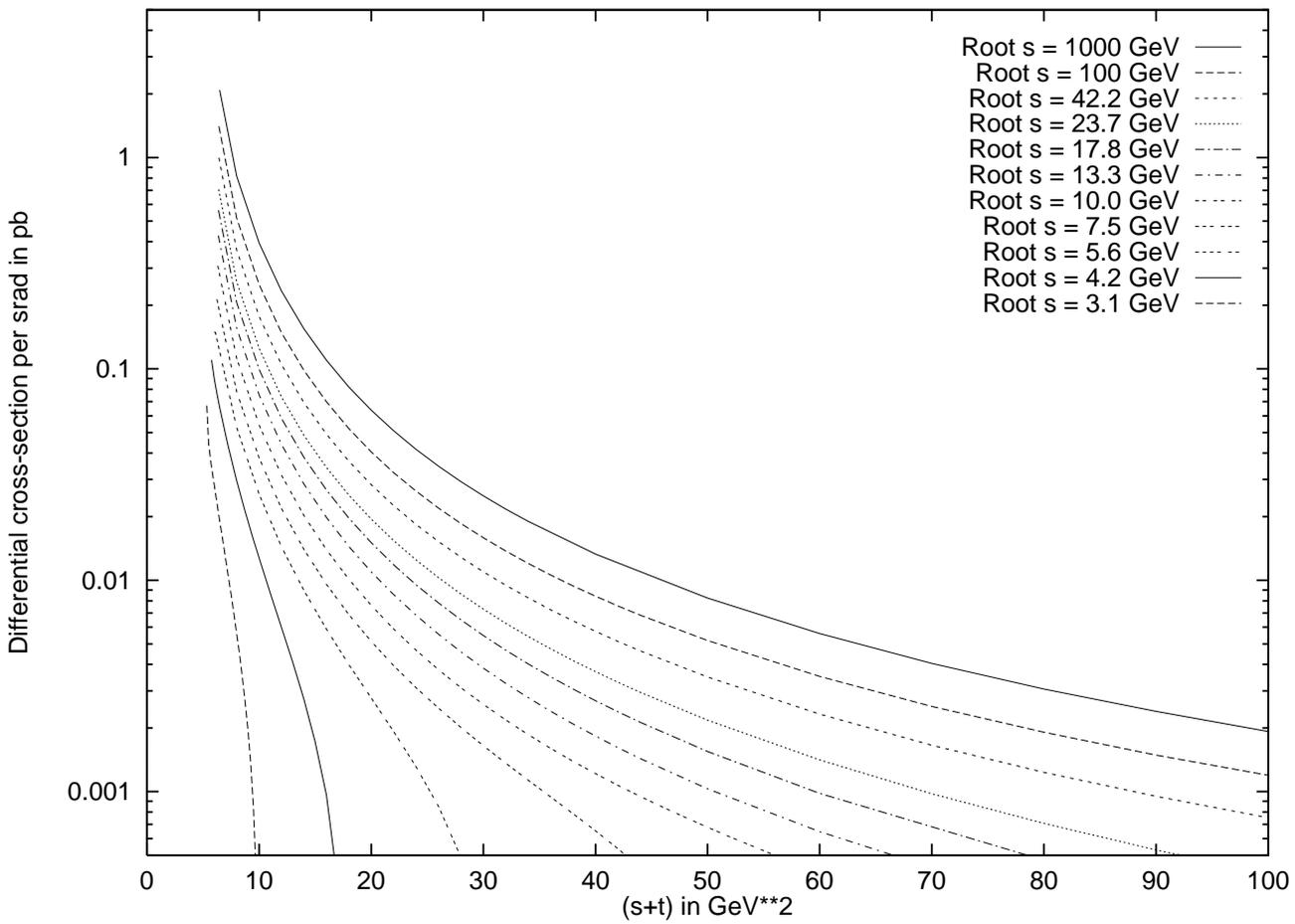,width=0.9\textwidth}}
\caption{Spin- and polarization-summed differential cross section $d\sigma/
   d\Omega$ for the transmutation process $\gamma e \longrightarrow \gamma 
   \tau$.}
\label{dc3}
\end{figure}
\begin{figure}
\centerline{\psfig{figure=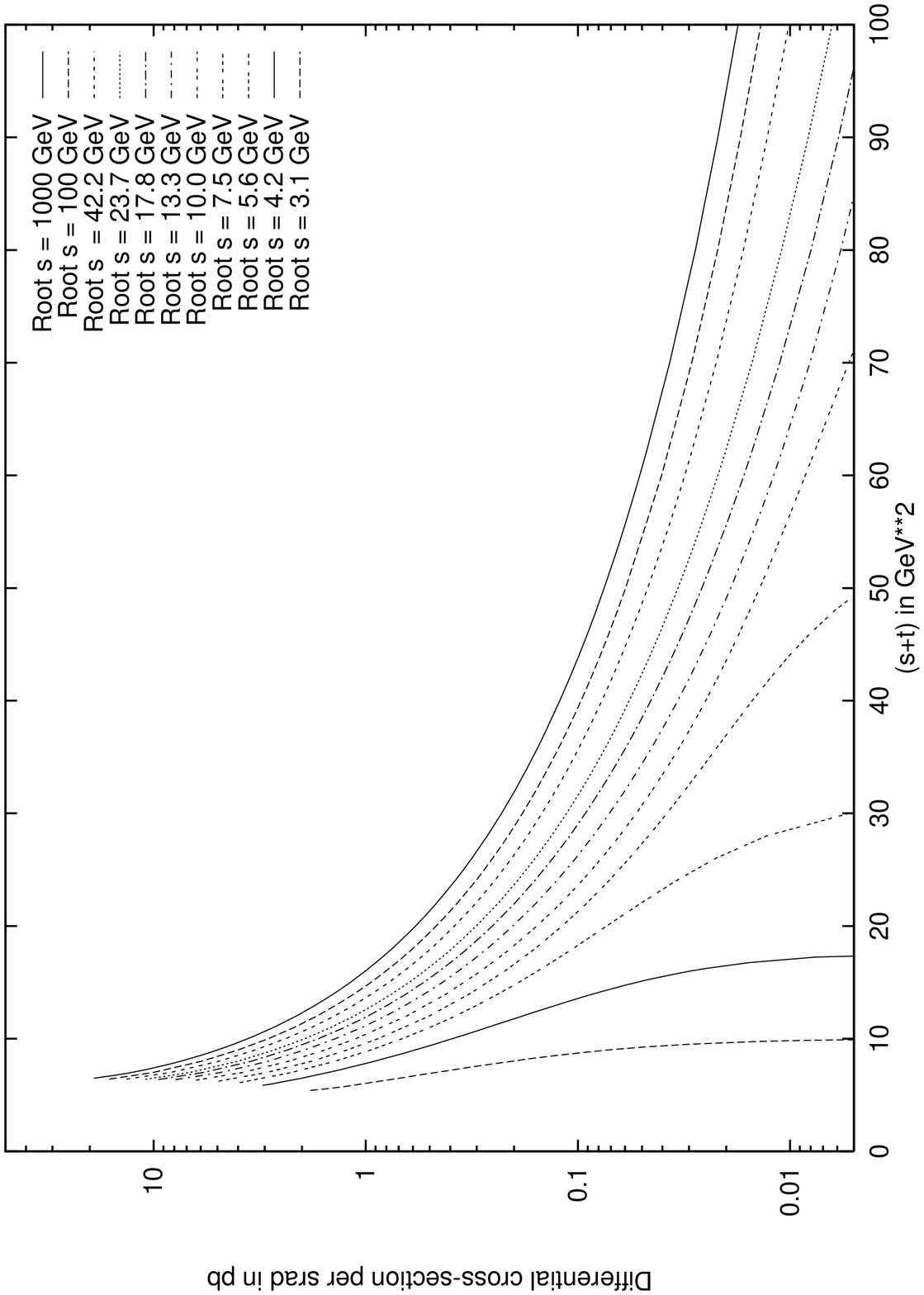,width=0.9\textwidth}}
\caption{Spin- and polarization-summed differential cross section $d\sigma/
   d\Omega$ for the transmutation process $\gamma \mu \longrightarrow \gamma 
   \tau$.}
\label{dc6}
\end{figure}
\begin{figure}
\centerline{\psfig{figure=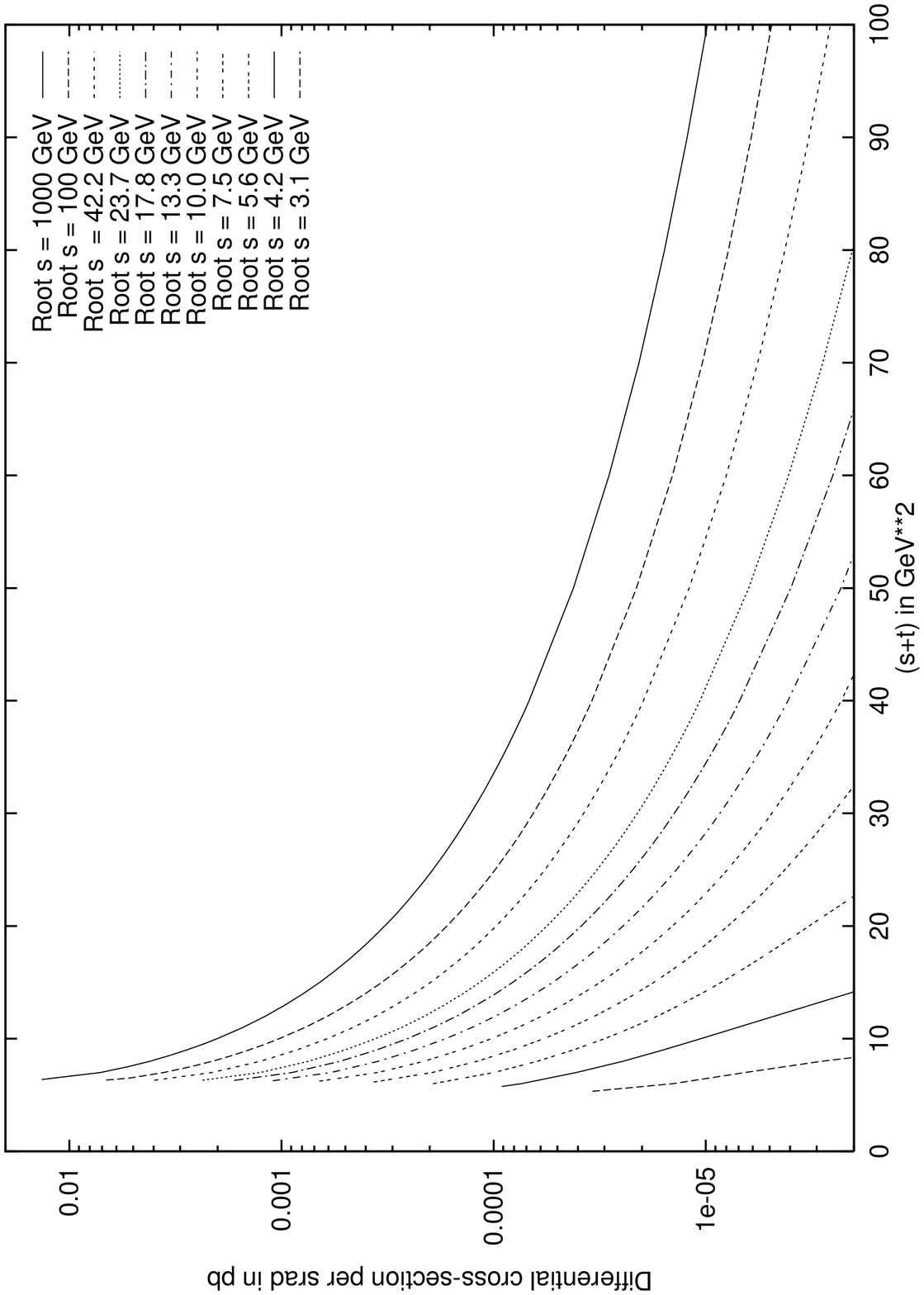,width=0.9\textwidth}}
\caption{Spin- and polarization-summed differential cross section $d\sigma/
   d\Omega$ for the transmutation process $\gamma e \longrightarrow \gamma 
   \mu$.}
\label{dc2}
\end{figure}
Above the $\tau$ threshold at $\sqrt{s} = 1.777$ GeV all 4 processes occur, 
the calculated cross sections for which are as presented in the Figures 
\ref{dc3}, \ref{dc6}, and \ref{dc2}.  They are seen to have very similar 
behaviour.  Indeed, the amplitudes for the different processes, according 
to (\ref{sumoveri}), differ at the same values of $s$ and $t$ only by the 
factors $S_{\alpha 1} S_{\beta 1}$, which depend only on the scale,
i.e.\  
$\sqrt{s}$, but not on $t$.  Hence, apart from small differences due to 
the kinematical factor in front in (\ref{diffxsec}), the plots in Figures
\ref{dc3}-\ref{dc2} at the same $s$ are nearly parallel for the different 
transmuatation processes, differing only by the factor $S_{\alpha 1}^2
S_{\beta 1}^2$.  In particular, given that the rotation matrix elements 
shown in Figure \ref{Salphai} are such that $|S_{\tau 1}| > |S_{\mu 1}|
> |S_{e1}|$, it follows that the processes are ordered in size of cross 
sections as follows: $\gamma \mu \longrightarrow \gamma \tau > \gamma e 
\longrightarrow \gamma \tau > \gamma e \longrightarrow \gamma \mu$.
Notice that, for ease of presentation, it is $d\sigma/d\Omega$ rather than 
the more usual $d\sigma/du$ that is plotted as a function of $s + t \sim u$ 
in Figures \ref{dc3}-\ref{dc2}.  This means that, in spite of appearances, 
the cross section is actually decreasing rapidly as $s \rightarrow \infty$.  
Notice further that in all the Figures \ref{dc3}-\ref{dc2}, the cross 
section is given only down to $s + t = 2 m_1^2 - m_1^4/s$ for $m_1 = m_\tau$ 
which is the smallest value of $s + t$ allowed within the physical region 
for the internal diagonal process $\gamma l_1 \longrightarrow \gamma l_1$.  
Smaller values of $s + t$ than these, down to $s + t = m_\alpha^2 + m_\beta^2
- m_\alpha^2 m_\beta^2/s$ are attained in the actual transmutation process 
$\gamma l_\alpha \longrightarrow \gamma l_\beta$ and for these values of 
$s + t$, the process $\gamma l_1 \longrightarrow \gamma l_1$ can no longer 
occur so that the transmutation amplitude is now by (\ref{sumoveri}) a 
sum over only the two remaining internal diagonal channels 2 and 3.  The 
plots in the Figures \ref{dc3}-\ref{dc2} of the transmutation cross sections 
should thus in principle be extended to smaller $s + t$ with values given 
by $2 S_{\alpha 3}^2 S_{\beta 3}^2$ times the Compton cross section, thus
showing a discontinuity at $s + t = 2 m_\tau^2 - m_\tau^4/s$.  
Such extensions are 
easily calculated but are awkward to present and therefore omitted in 
our figures.  We are in fact uncertain whether the noted discontinuity 
at $s + t = 2 m_\tau^2 - m_\tau^4/s$ will in fact appear in the
physical cross section
or whether it will be smoothed out by other effects not yet considered.

At energies $\sqrt{s}$ below the $\tau$ mass, only the transmutations
$\gamma e \longrightarrow \gamma \mu$ and $\gamma \mu \longrightarrow 
\gamma e$ can occur with approximately equal cross sections.  The result
of our calculation for $\gamma e \longrightarrow \gamma \mu$ is shown 
in Figure \ref{dc2lowe}.  
\begin{figure}
\centerline{\psfig{figure=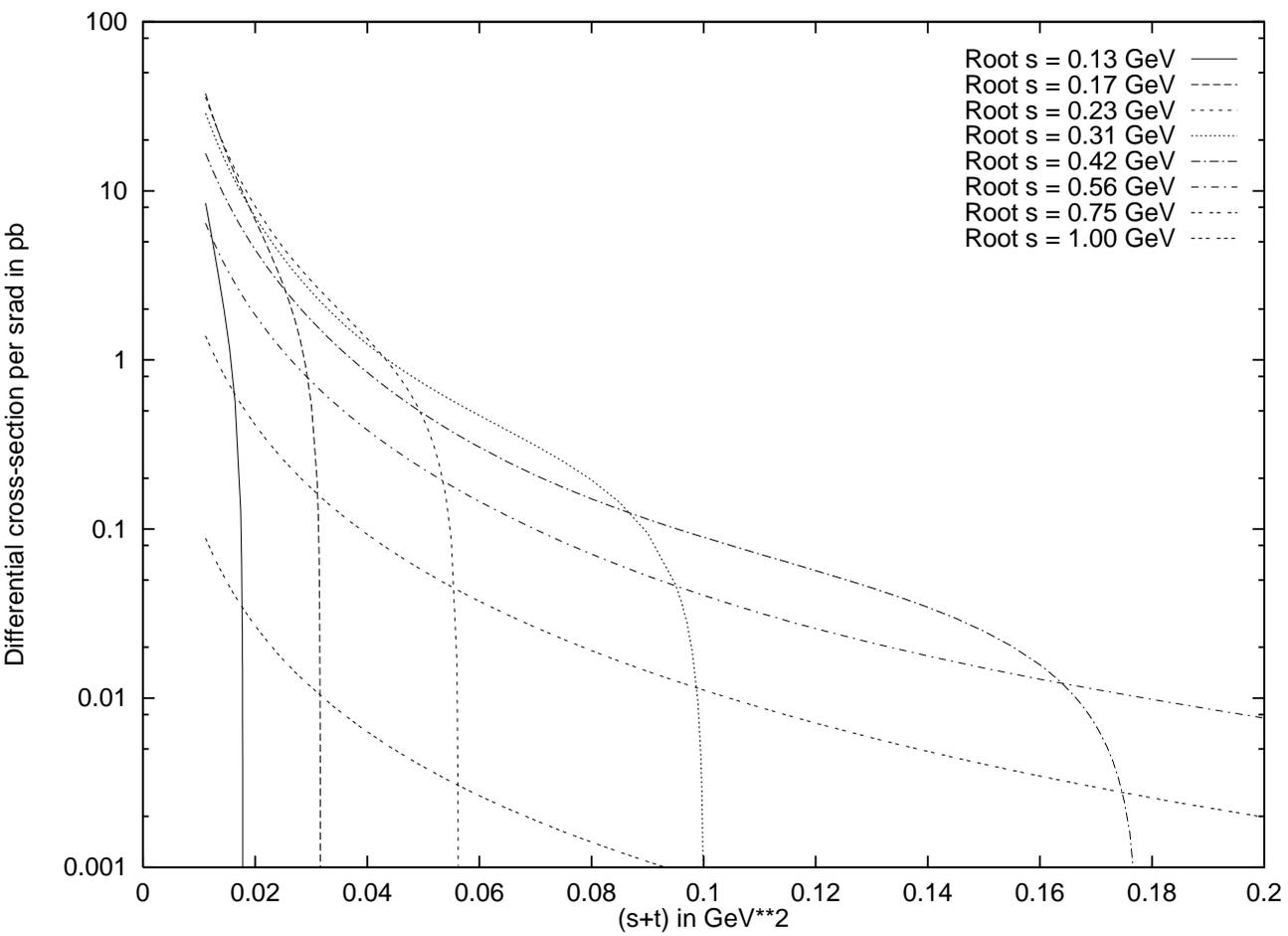,width=0.9\textwidth}}
\caption{Spin- and polarization-summed differential cross section $d\sigma/
   d\Omega$ for the transmutation process $\gamma e \longrightarrow \gamma 
   \mu$ below $\tau$ threshold.}
\label{dc2lowe}
\end{figure}
Though still dominated by the $u$-channel pole, 
the cross section has here a more intricate behaviour than at the higher 
energies shown in Figure \ref{dc2}.  First, the mass matrix being by 
construction in SSD diagonal when the scale equals either the $\mu$ or 
the $\tau$ mass, the transmutation cross section is vanishing at both 
these energies.  It thus shows first a rise from zero at the threshold 
$\sqrt{s} = m_\mu$ of the reaction to a maximum shortly above, and then 
falls to zero again at $\sqrt{s} = m_\tau$.  Secondly, although the
mass matrix here has again just one nonzero eigenvalue as for $\sqrt{s}$ 
above the $\tau$ mass, this nonzero eigenvalue $m_2$ itself approaches 
zero as $\sqrt{s} \rightarrow m_\tau$.  In fact, as can be seen in 
Figure \ref{Salphai}, $m_2$ remains rather small all the way until 
$\sqrt{s}$ approaches $m_\mu$, where of course it equals $m_\mu$ by 
construction.  Hence, in contrast to the high $s$ situation in Figure 
\ref{dc3}-\ref{dc2}, the smallest value of $s + t$, namely $2 m_2^2 - 
m_2^4/s$, within the physical region of the internal process $\gamma l_2 
\longrightarrow \gamma l_2$, is in this case usually smaller than the 
smallest value of $s + t$, namely $m_e^2 + m_\mu^2 - m_e^2 m_\mu^2/s$, 
within the physical region of the actual transmutation process $\gamma e 
\longrightarrow \gamma \mu$ under consideration.  For this reason, in 
Figure \ref{dc2lowe}, the cross section is given down to $s + t = m_e^2 
+ m_\mu^2 - m_e^2 m_\mu^2/s$ without showing any discontinuity in $s + t$, 
although, in principle, at energies where this limit can become smaller 
than $s + t = 2 m_2^2 - m_2^4/s$, similar remarks to those given at the 
end of the last paragraph also apply.  By numerically integrating the
differential cross section given in Figure \ref{dc2lowe}, one obtains 
approximately the total cross section for the process $\gamma e 
\longrightarrow \gamma \mu$ shown in Figure \ref{totalc}, which shows 
an intriguing maximum at cm energy of about 180 MeV.  
\begin{figure}
\centering
\includegraphics[angle=-90,scale=0.75]{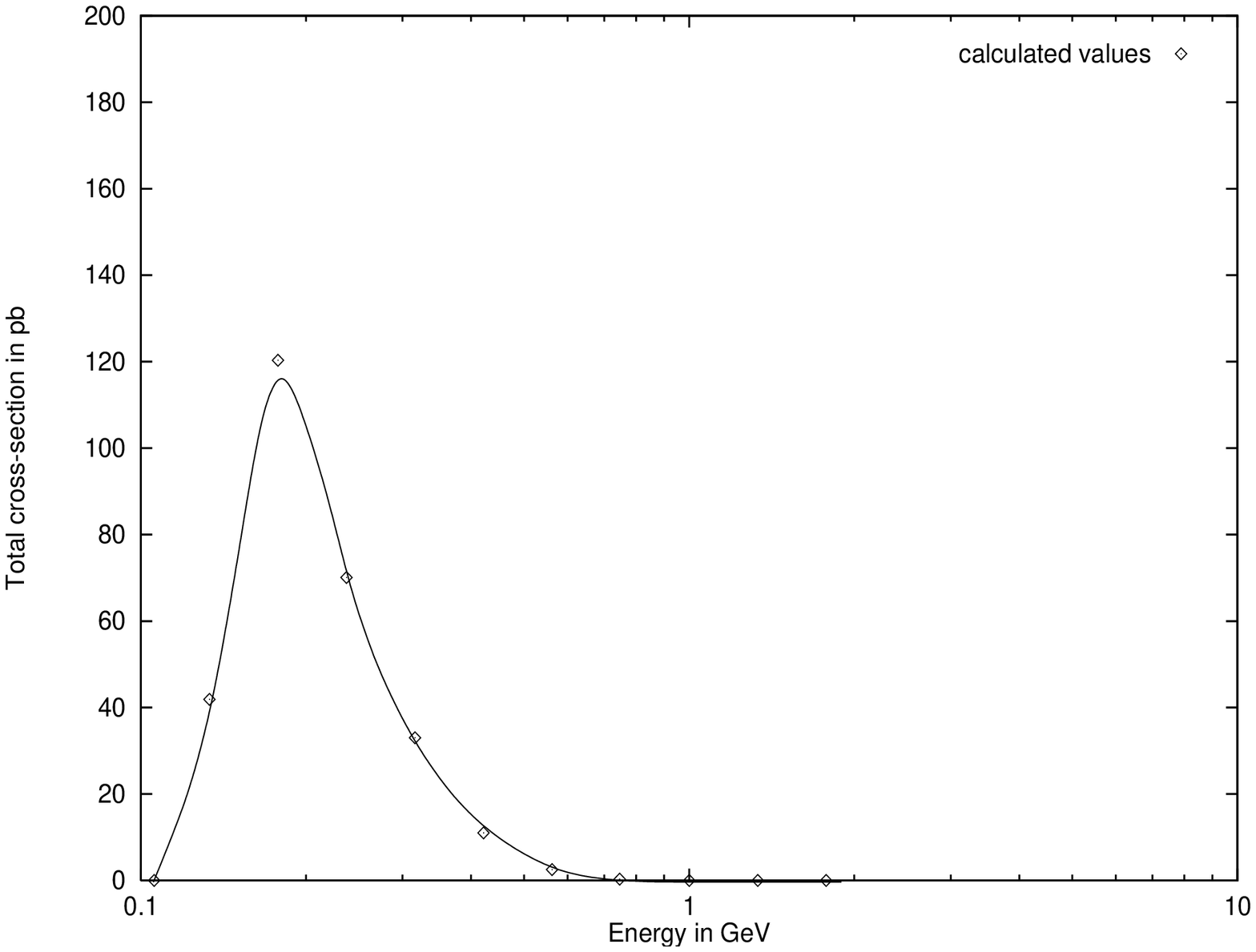}
\caption{Total cross section for the transmutation process $\gamma e
   \longrightarrow \gamma \mu$ below the $\tau$ threshold.}
\label{totalc}
\end{figure}

Although the above calculation has been performed in detail only for the 
DSM scheme, from this the transmutation cross sections for the NSM, 
or indeed any other rotation, scheme 
can also readily be inferred.  In general, of course, the mass matrix 
has no reason to have two zero eigenvalues as in the DSM 
case, but the mass eigenvalues must still be hierarchical, meaning that 
$m_1 \gg m_2 \gg m_3$.  In such a case, given that $m_2 - m_3$ is necessarily 
much smaller than $m_1 - m_2$ or $m_1- m_3$, we can neglect in the amplitudes 
the difference ${\cal M}_{2} - {\cal M}_{3}$ compared with the difference 
${\cal M}_{1} - {\cal M}_{3}$ so that the expression in (\ref{sumoveri}) 
remains approximately valid.  Then at energies above the $\tau$ mass, the 
mass $m_1$ being necessarily close to $m_\tau$, the cross section will 
be approximately given just by re-scaling the cross section for DSM at the 
same value of $s + t$ by a factor $[S_{\alpha 1}^2 S_{\beta 1}^2]_{any}/ 
[S_{\alpha 1}^2 S_{\beta 1}^2]_{DSM}$.  For example, for the
transmutation $\gamma \mu \longrightarrow \gamma \tau$ in NSM from
equation (4.2) of \cite{impromat} one obtains $S_{\mu 1} S_{\tau 1}
\sim \langle \mu | m | \tau \rangle /m_1$.  At $\sqrt{s} = 17.8$ GeV,
say, one decade above the $\tau$ mass when the mass matrix is
diagonal, $S_{\mu 1} S_{\tau 1} \sim 3 \times 10^{-3}$ compared with
$\sim 5 \times 10^{-2}$ for DSM as read from Figure \ref{Salphai}
here.  Hence the differential cross section at this energy for NSM is
just given by multiplying the curve at the same energy in Figure
\ref{dc6} of this paper by a factor $\sim (0.06)^2$, i.e.\ some 2--3
orders smaller.  At scales below the $\tau$ mass, this simple
criterion no longer works, but here in any case
the NSM scheme has presumably to be modified in view of the
apparent difficulty with the ``pole in the wrong channel'' mentioned
above, so that one is yet in no position to calculate the transmutation
cross section.

\setcounter{equation}{0}

\section{Remarks}

Our first aim in this paper is to examine the possibility of, and if
possible to develop the methodology for, calculating cross sections of
transmutation processes when the fermion mass matrix rotates with 
changing scales, using photo-transmuation as a specific example.  One 
ends up with a procedure which seems both to be internally consistent and 
to give reasonable answers, while offering a chance for generalization
to other processes.  The procedure was constructed following a certain 
line of logic, although one cannot pretend in any sense to have derived 
it from first princples.  One can thus hope that it will serve at least 
as a working hypothesis for exploring the unfamiliar physics involved.  
Even with the method so developed, the calculation still depends on the 
physics mechanism driving the rotation of the mass matrix, and the result 
one obtains, of course, can at best be as good as the physics mechanism 
from which it is deduced.  We have investigated two particular schemes 
labelled respectively as NSM and DSM.  Although we are biased towards 
the second, this being the scheme we ourselves suggested and advocate, 
we are aware of some shortcomings \cite{dsmrph}
and its need for further refinement.  
The results in either scheme therefore have yet to be taken with caution.

With these reservations in mind, let us proceed to analyse the results 
obtained.  We have not as yet made a thorough study as regards their
accessiblity to tests by experiment.  Some general observations, however,
are already possible.  In looking for photo-transmutation effects, one
need not restrict oneself to real photons but can make free use of the
virtual photons available in $e^+ e^-$ colliders.\footnote{We are grateful 
to John Guy for a reminder and subsequent discussion for this possibility.}.  
However, the cross sections at energies above the $\tau$ mass being in 
general rather small, namely at most in the low picobarn or multi-femtobarn 
range shortly above the $\tau$ threshold, and decreasing rapidly with
increasing energy, it seems unlikely that they would have been observed 
without a conscious search for them.  This is true for both the NSM and
DSM schemes.  To ascertain whether some of the effects would be observable 
in existing machines such as LEP or BEPC, some closer study would be 
necessary, which is underway.  The cross section being much larger for 
transmutation of $e$ into $\tau$ than of $e$ into $\mu$, as can be seen 
in Figures \ref{dc3} and \ref{dc2} for DSM, a search for $\tau$ seems 
indicated.  In the future, when $\mu$ storage rings become available
then a search for the transmutation for $\mu$ into $\tau$ looks even
better because of the larger cross section, as seen in Figure \ref{dc6}
for DSM.  The same remarks should hold also for NSM.

To look for $e$ to $\mu$ transmutation, it would be much more profitable 
to work at low energy below the $\tau$ threshold at cm energy of around 
200 MeV, where the cross section in DSM has a maximum, as can be seen in 
Figure \ref{totalc}.  At first sight, the predicted cross section being 
of the order of 100 picobarns, it seems surprising that the effect have 
not been seen already by experiment, but we have not yet succeeded 
in locating any relevant data on $\gamma e$ collisions in that energy 
region.  The reason seems to be that with electrons having a mass of 
only half an MeV, it will take 40 GeV photons impinging on a stationary 
electron target to make up a cm energy of 200 MeV, and real photon beams 
of such an energy are not readily available.  One can hope to make up
the desired energy also by shining laser light on to one of the two beams
in a electron-positron collider, or else the back-scattered light from
one beam on to the other, but none of the existing colliders we looked
at seem to have just the right beam energies for this purpose, although
it would not be difficult of course to achieve this energy for a 
purpose-built collider.  Otherwise, one can utilize again the virtual
photons available in $e^+ e^-$ colliders, in which case low energy
colliders like BEPC would seem to be more appropriate.

As far as searching for transmutational effects is concerned, there is no 
reason of course to restrict to photo-transmutations.  The same general 
considerations can be applied to other processes, e.g. to $e^+ e^-$ collison
itself giving transmutations like $e^+ e^- \longrightarrow e^+ \mu^-$.
The cross sections for these processes have not yet been calculated but
from the experience gained in this paper, one can hope that the methodology
developed here can be adapted to $e^+ e^-$ collision with only moderate 
modifications.  If so, then our result in Figure \ref{totalc} for 
photo-transmutation suggests that there may also be a maximum in the cross
section for $e^+ e^- \longrightarrow e^+ \mu^-$ at cm energy of a few 
hundred MeV.

In any case, there can be a wide field here for future experimental 
investigation.


\begin{thebibliography}{99}

\bibitem{rge} See e.g. B. Grzadkowski, M. Lindner and S. Theisen,
   Phys. Lett. B198, 64, (1987); H.\ Arason, D.J.\ Casta\~no, B.\ 
   Kesthelyi, S.\ Mikaelian, E.J.\ Piard, P.\ Ramond and B.D.\ Wright, 
   Phys.\ Rev.\ D46 (1992) 3945.

\bibitem{impromat} Jos\'e Bordes, Chan Hong-Mo and Tsou Sheung Tsun,
   hep-ph/0006338.

\bibitem{ckm} N.\ Cabibbo, Phys.\ Rev.\ Lett.\ 10, 531 (1963); 
   M.\ Kobayashi and T.\ Maskawa, Prog.\ Teor.\ Phys.\ 49, 652
   (1973).

\bibitem{mns} Z.\ Maki, M.\ Nakagawa and S. Sakata, Prog.\ Theor.\ Phys. 
   28 (1962) 247.

\bibitem{soudan} T. Kafka, Nucl. Phys. B (Proc. Suppl.) 35, 427, (1994);
   M. Goodman, {\it ibid} 38, 337, (1995); W.W.M. Allison et al., Phys. 
   Letters B391, 491, (1997).

\bibitem{chooz}  CHOOZ collaboration, M. Apollonio et al., Phys. Lett.
   B420, 397, (1997).

\bibitem{superk} Y.\ Fukuda et al, Super-Kamiokande Collaboration, 
   Phys.\ Lett.\ B433 (1998) 9; Phys.\ Lett.\ B436 (1998) 
   33; Phys.\ Rev.\ Lett.\ 81 (1998) 1562.

\bibitem{physcons} Chan Hong-Mo and Tsou Sheung Tsun, Phys.\ Rev.\ 57D,
   (1998) 2507, hep-th/9701120.

\bibitem{ourckm} Jos\'e Bordes, Chan Hong-Mo, Jacqueline Faridani, Jakov
   Pfaudler, and Tsou Sheung Tsun, Phys.\ Rev.\ D58, 013004, (1998),
   hep-ph/9712276.

\bibitem{ournuos} Jos\'e Bordes, Chan Hong-Mo, Jakov Pfaulder and Tsou
   Sheung Tsun, Phys.\ Rev.\ D58 (1998) 053003, hep-ph/9802420;
   Phys.\ Rev.\ D58 (1998) 053006, hep-ph/9802436.

\bibitem{phenodsm} Jos\'e Bordes, Chan Hong-Mo and Tsou Sheung Tsun,
   hep-ph/9901440, Eur. Phys.\ J.\ C10 (1999) 63.

\bibitem{dualgen} Chan Hong-Mo, to appear in the proceedings of {\em 
   Intern.\ Conf.\ on Fundamental Sciences: Mathematics and
   Theoretical Physics}, March 2000, Singapore.

\bibitem{Mandshaw} F.\ Mandl and G.\ Shaw, {\em Quantum Field Theory},
   John Wiley and Sons, 1984. 

\bibitem{dsmrph} Chan Hong-Mo and Tsou Sheung Tsun, Int.\ J.\ Mod.\
   Phys.\ A14 (1999) 2173, hep-ph/9904406.

\end{thebibliography}
\end{document}